\documentstyle[epsf,12pt]{article}

\newcommand{\eq}[1]{eq.~(\ref{#1})}
\newcommand{\eqs}[2]{eqs.~(\ref{#1},\ref{#2})}

\newcommand{\ur}[1]{(\ref{#1})}

\newcommand{\urss}[3]{~(\ref{#1},~\ref{#2},~\ref{#3})}

\newcommand{\fig}[1]{Fig.~\ref{#1}}

\newcommand{\beq}{\begin{equation}}
\newcommand{\eeq}{\end{equation}}

\newcommand{\bI}{\bar I}

\newcommand{\la}[1]{\label{#1}}

\newcommand{\beqa}{\begin{eqnarray}}
\newcommand{\eeqa}{\end{eqnarray}}
\newcommand{\Tr}[1]{{\rm Tr}\left[#1\right]}

\def\pj{\hspace{-.2cm}}
\newcommand\dbar[1]{ \frac{d^4#1}{(2\pi)^4}}

\def\Dirac#1{#1\hskip-6pt/}
\def\dd{\Dirac\partial}

\begin{document}
\begin{flushright}{\normalsize NORDITA--98/68\\
NBI--98--39\\
December 1998\\}
\end{flushright}
\vskip 2true cm
\begin{center}
{\Large\bf Light Quarks in the Instanton Vacuum \\
\vskip .5true cm
at Finite Baryon Density}

\vskip 1true cm
{G.W. Carter}\\
\vskip .5true cm
{\it Niels Bohr Institute, Blegdamsvej 17, DK--2100 Copenhagen,
Denmark} \\
\vskip .5true cm
{D. Diakonov} \\
\vskip .5true cm
{\it NORDITA, Blegdamsvej 17, DK--2100 Copenhagen, Denmark} \\
\date{}

\parindent=20pt
\end{center}

\vskip 2true cm
\begin{center}
{\bf Abstract}\\
\end{center}

\noindent
We consider the finite density, zero-temperature behaviour of quark matter
in the instanton picture. Since the instanton-induced interactions are
attractive in both $\bar{q}q$ and $qq$ channels, a competition ensues
between phases of matter with condensation in either or both.
It results in chiral symmetry restoration due to the onset of diquark
condensation, a `colour supercondutor', at finite density.  Also present
is a state with both manners of condensation, however this phase
is found to be thermodynamically disfavoured for equilibrium matter.
Properties of quark matter in each phase are
discussed, with emphasis on the microscopic effects of the effective
mass and superconducting energy gap.
\vskip 1true cm \noindent

\thispagestyle{empty}

\vfill \eject

\section{Introduction}

The idea that the QCD partition function is dominated
by instanton fluctuations of the gluon field, with quantum
oscillations about them, has successfully confronted the majority of
facts we know about the zero-temperature, zero-density hadronic world
(for reviews see refs.\cite{D1,SS}).  Instantons have been reliably
identified in lattice simulations (for a review see ref.\cite{vB}), and
their relevance to hadronic observables clearly demonstrated
\cite{Negele}. From the theory side, the instanton vacuum constructed
from the Feynman variational principle \cite{DP1,DPW}, gives 
an example of how the necessary `transmutation of dimensions' can
actually happen in QCD, meaning that all dimensional quantities
can be expressed through the QCD scale parameter, $\Lambda_{QCD}$. 
It can be added that in the solvable $N=2$ supersymmetric version of QCD 
it is instantons --- and nothing besides them --- that are sufficient 
to reproduce the expansion of the exact Seiberg--Witten prepotential 
\cite{DKM}.

Of even more relevance to the topic of this paper, the instanton vacuum
provides a theoretically beautiful and phenomenologically
realistic mechanism of spontaneous chiral symmetry breaking in QCD
(for a review see \cite{D1}). In addition, it has recently
gained a strong support from direct lattice studies \cite{deF}.
Therefore an expansion of instanton ideas to a new frontier,
in this case to nonzero matter density, seems to be well justified.

The key point is that instantons induce interactions that are
attractive not only in the $\bar q q$ channel (leading to the
spontaneous chiral symmetry breaking) but also in the $qq$ channel
(potentially leading to diquark condensation). This has been first
noticed in ref. \cite{BeLa}. It becomes especially clear in the
special case of two colours ($N_c=2$) and two light quark flavours
($N_f=2$). In this case instanton-induced interactions possess
not a global $SU(2)\times SU(2)$ as would be for any $N_c>2$, but
a larger $SU(4)$ symmetry \cite{DP5} (often referred to as
Pauli--G\"ursey symmetry). At $N_c=2$ the attraction in the $\bar qq$
and $qq$ channels are exactly equal. For this reason $\bar q q$ and
$qq$ condensates belong in fact to one phase: one condensate can be
rotated to another along the Goldstone valley \cite{DP5}\footnote{In
the original paper \cite{DP5} a general case of the $SU(4)$ breaking
has been considered, leading to nine Goldstone particles. A closer
inspection shows, however, that the instanton-induced interactions
possess an additional degeneracy, leading to only five Goldstone
particles, instead of nine which would be the general case. Instantons
apparently `know' about the Vafa--Witten theorem which guarantees that
the symmetry breaking pattern at $N_c=2$ corresponds to five Goldstone
particles \cite{SV}.}.

Switching in a nonzero chemical potential $\mu$ violates explicitly
the global $SU(4)$ symmetry of the $N_c=2$ world, the degeneracy of the
5-dimensional Goldstone valey is lifted, and one can ask which
of the condensates becomes preferred. In ref. \cite{StonyBrook} arguments
have been given that it is the diquark condensate. In a previous
publication \cite{CD} we have confirmed this expectation by direct
calculations; see also below. It should be noted that at $N_c=2$
singlet diquarks are nothing but the colourless `baryons' which
happen to be bosons, and their condensation does not break the
colour symmetry.

In the opposite limit, $N_c\to\infty$, the diquark interaction
is suppressed by a factor of $\sim 1/N_c$ as compared to the
$\bar qq$ one, and so is the diquark condensate. The question
is then, is our world with $N_c=3$ closer to $N_c=2$ or
$N_c=\infty$?

The possibility of diquark condensation at any $N_c$, as induced by
instantons, has been studied in ref. \cite{DFL}. Contrary to the case
of $N_c=2$, a diquark condensation at $N_c\geq 3$ would inevitably
spontaneously break the colour symmetry, similarly to the Higgs
breaking of the $SU(2)$ gauge symmetry of electroweak interactions.
For that reason the (possible) symmetry breaking of the colour
$SU(N_c)$ has been named `dynamical Higgs mechanism' in ref.\cite{DFL},
and a parallel with the superconductivity has been drawn. The presently
used term is `colour superconductivity' to which we shall adhere.

For zero chemical potential only a metastable
diquark-condensed state has been found at $N_c\geq 3$ \cite{DFL}.
Furthermore, at $N_c=3$ the scalar diquark already appears to be
unbound in the vacuum, indicating that our world is in a sense closer
to the idealized $N_c\to\infty$ than to the $N_c=2$ world.
Parametrically, the diquark mass is $\sim 1/\bar\rho\sim 1\;{\rm GeV}$
where $\bar\rho$ is the average instanton size as explained below. This
means that a scalar diquark correlation function should decay with the
exponent corresponding to the `constituent' quark threshold
$2M(0)\approx 700\;{\rm MeV}$, which seems to be supported by recent
lattice measurements \cite{HKLW}.  Nevertheless, it has been suggested
in \cite{DFL} that $qq$ condensates could be found as metastable states
in heavy ion collisions and in astrophysics.

More recently it has been argued by the Princeton/MIT and Stony Brook
groups \cite{StonyBrook,Princeton/MIT} both using instantons as a
framework, that taking nonzero fermion density shifts the balance in
favour of the diquark condensation, and that at certain critical
chemical potential $\mu_c$ there should be a phase transition from the
usual broken chiral phase to the colour superconducting state\footnote{
See the Proceedings of the International Workshop of
QCD at Finite Baryon Density (Bielefeld, Germany), where
these matters have been extensively discussed \cite{Biel}.}. Avoiding
some of the unnecessary approximations made in
\cite{StonyBrook,Princeton/MIT} we arrive essentially at the same
conclusions \cite{CD}.

In this paper we study the competition between various quark channels
in a more systematic way than in our previous publication \cite{CD}.
Since `colour superconductivity' implies colour symmetry is broken,
one can imagine several phases with chiral symmetry broken or restored
for quarks of different colours. We explore these possibilities using
what amounts to a virial expansion in the instanton density, and
carrying out detailed calculations to first order.

The chiral broken phase is characterized by a nonzero order parameter
$\langle\bar q q\rangle\neq 0$ or, equivalently, by a nonzero dynamical
or `constituent' quark mass which in fact is momentum-dependent. We
find that, as far as one remains inside this phase, the quark
occupation number $n(p)$ is a perfect Fermi step function, despite
strong energy and momentum-dependent interactions between quarks. It should be
contrasted to the colour superconducting phase with
$\langle qq\rangle \neq 0$ but $\langle\bar q q\rangle = 0$.
In this case the quark occupation number $n(p)$ is distorted near the
Fermi surface and is of a typical BCS type.

In our numerics we use the standard characteristics of the instanton
ensemble. They lead to very reasonable values of the quark condensate
and of the constituent quark mass at zero chemical potential. However,
the same instanton characteristics lead inevitably to a very early
transition to the superconducting state: it occurs at quark densities
less than the normal nuclear density; after the system jumps into
the superconducting phase the density appears to be about twice the
nuclear density. Taken literally, it suggests that the
ordinary nuclear matter is a `boiling' mixture of the two phases
\cite{BR}, however, it most probably contradicts the lore of
conventional nuclear theory\footnote{We take an
opportunity to thank G.Brown and K.Langanke
for a instructive discussion of this point.}. To `save' the nuclear
matter, one would probably need to go beyond the mean-field
approximation actually used in this paper.

\section{QCD Instanton Vacuum}

The construction the QCD instanton vacuum has been
reviewed in previous publications \cite{D1,CD}, therefore here
we simply review the main steps.

\begin{itemize}
\item  The use of the Feynman variational principle applied to
instantons leads to the stabilization of the grand canonical
instanton ensemble, with the main characteristics of the instanton
medium, namely, the average 4-dimensional instanton density, $N/V=1/\bar R^4$,
and the average instanton size, $\bar \rho$, expressed through
the only dimensional quantity $\Lambda_{QCD}$. The 2-loop calculations
performed in \cite{DP1,DPW} give

\beq \bar\rho =
\sqrt{\overline{\rho^2}} \approx \frac{0.48}{\Lambda_{\overline{MS}}},
\;\;\;\;\; \bar R \approx \frac{1.35}{\Lambda_{\overline{MS}}}.
\la{rhoR}\eeq
Taking $\Lambda_{\overline{MS}}=280\;{\rm MeV}$ one finds
$\bar\rho\approx 0.35\;{\rm fm},\;\;\bar R \approx 0.95\:{\rm fm},
\;\;\bar\rho/\bar R\approx 1/3$. This small ratio has been previously
suggested on phenomenological grounds by Shuryak \cite{Sh1}.
The smallness of the $\bar\rho/\bar R$ ratio implies that
the packing fraction of instantons in the vacuum, i.e. the
fraction of the 4d volume occupied by the balls of radius
$\bar\rho$, is quite small:
\beq
f=\frac{\pi^2}{2}\bar\rho^4\frac{N}{V}
=\frac{\pi^2}{2}\frac{\bar\rho^4}{\bar R^4}\sim\frac{1}{10}.
\la{pack}\eeq

\item
Though the packing fraction \ur{pack} is but numerically small, one
can treat it as a formal algebraic parameter, and develop a
perturbation theory in it. Taking nonzero matter density and/or
nonzero temperature can only decrease the ratio $\bar\rho/\bar R$.

\item
When one switches in light quarks on top of the instanton ensemble,
the existence of the small packing fraction parameter enables one
to separate contributions of high and low fermion eigenmodes. The
high-frequency part is controllably factorizable into contributions
from individual instantons. This part can be seen to renormalize
somewhat the one-instanton weight or instanton `fugacity'. The
low-frequency part is dominated by the would-be zero modes of
individual instantons. The normalized (that is localized) zero modes
in the background of individual instantons exist both in the
vacuum ($\mu=0$) \cite{tH} and at $\mu\neq 0$ \cite{Abr,DeCarv},
see the Appendix. In the ensemble of instantons and anti-instantons
they cease to be exactly zero modes (that is why we call them the
`would-be zero modes'). The spectral density of the Dirac operator
at small eigenvalues is obtained through the diagonalization of the
matrix made of the overlaps of the would-be zero modes belonging to
individual instantons and anti-instantons \cite{DP2}. A nonzero
spectral density at zero Dirac eigenvalue signals chiral symmetry
breaking: it is actually due to the delocalization of the would-be zero
modes owing to the `hopping' of quarks from one instanton to another
\cite{DP3}.

\item
In fact, three seemingly different but actually equivalent methods have
been developed to describe chiral symmetry breaking by instantons at
zero $\mu$: {\it i)} diagonalization of the random matrix made of
the would-be zero modes' overlaps \cite{DP2,Sim}; {\it ii)} finding
the quark propagator in the random instanton ensemble \cite{DP3,Pob},
which exhibits the appearance of a momentum-dependent quark mass;
{\it iii)} derivation of the effective low-momentum theory for quarks
with instanton-induced interactions \cite{D1,DP4}.
\end{itemize}

The three approaches underline different sides of the physics involved
though mathematically they prove to be equivalent. It is
straightforward to generalize them to the case of nonzero chemical
potential. In this paper we use only the third method, it being the most
economical.

\section{Effective Fermion Action for Finite Density}

With the instanton solution representing a dominant, stable fluctuation of
the gauge field, the original strategy for computing quark observables in
the QCD vacuum was to calculate the quantity in the presence of a random
instanton background and then average over
an ensemble of such field configurations.  However, the analysis here calls for
a more transparent manifestation of the effects from instantons in order to
resolve the possible mechanisms for symmetry breaking.  This is obtained by
first averaging over the ensemble of instanton and anti-instanton
configurations to formulate an effective theory in terms of interacting
quarks, where explicit instanton effects are absorbed into the form of the
interaction.

This evidence of averaged instantons is retained in the interaction through
the would-be fermion zero modes, one for each flavour.  Thus
the consequent interaction is a vertex involving $2N_f$ quarks, commonly
cited as a 't Hooft interaction after the first author to specify their
proper quantum numbers \cite{tH}.  For two flavours, it may be cast to
resemble the Vaks-Larkin/Nambu-Jona-Lasinio model which has long been
a popular model for quark and hadron phenomenology.

The field of a single instanton/anti-instanton determines a particular
solution for the gauge field, $A^{\mu}_I$, and in the absence of quantum
fluctuations the zero mode, $\Phi(x,\mu)$, is simply the solution of the
Dirac equation with zero eigenvalue:
\beq
\left( i\dd - i\mu\gamma_4 - {\Dirac A}_I \right)\Phi(x,\mu) =
 0\cdot\Phi(x,\mu)=0 \,.
\label{diraceq}
\eeq
The exact solution
for nonzero $\mu$ was found by Abrikosov \cite{Abr} and is written in
the Appendix.  It can be decomposed into chiral compontents as
\beq
\Phi(x,\mu) = \left[\begin{array}{c} \Phi_{\bI}(x,\mu) \\
\Phi_I(x,\mu) \end{array}\right] \,,
\eeq
meaning that a (right) left handed zero mode is generated by an
(anti-) instanton.  This is a result of the instanton solution's
structure and the self-dual equations from which it comes.
With $\tilde \Phi_{I(\bI)}$ we will denote the conjugate zero mode, which
takes the chemical potential argument of opposite sign:
$\tilde \Phi_I(x,\mu)=\Phi_I^\dagger(x,-\mu)$, where
the dagger means hermitean conjugate.  This is a direct consequence of
the non-hermiticity of the $\mu$-dependence which arises in the Dirac
operator.

This solution is not physically realizable, since the QCD background
is not modelled by one but many (a `liquid' of) instantons.
Localized around each there is but a would-be zero mode, from which the
effective action is constructed.  This action is determined by building a
low-momentum partition function, the foundation of which is
the Green function for a quark in the field of one instanton.
It can be expressed as a sum over the complete set of eigenfunctions, and
approximated as
\beqa
S^I(x,y) &\pj\equiv&\pj \langle \psi(x)\psi^\dagger(y)\rangle =
- \sum_n \frac{\psi_n(x)\psi^\dagger_n(y)}{\lambda_n+im} \nonumber\\
&\pj\approx&\pj -\frac{\Phi(x)\tilde{\Phi}(y)}{im} + S_0(x,y;\mu) \,.
\label{splitgf}
\eeqa
At low momenta this propagator is dominated by the zero modes while
at large momenta it is essentially reduced to the free one, 
$S_0(x,y;\mu)$. Therefore, \eq{splitgf} can be regarded as an 
interpolation of the true propagator in the field of one instanton. 
It becomes exact both at large, $p\gg 1/\bar\rho$, and small, 
$p\ll 1/\bar\rho$, quark momenta.

The partition function which produces the necessary propagator (\ref{splitgf})
is of the form
\beq {\cal Z}
=\int\!D\psi\:D\psi^\dagger\:\exp\left[\sum_f\!\int\!
\psi_f^\dagger(i\dd-i\mu\gamma_4)\psi_f\right]
\left(\frac{Y^+_{N_f}}{V} \right)^{N_+} \left( \frac{Y^-_{N_f}}{V}
\right)^{N_-}.
\la{Z4}\eeq
The pre-exponential factors contain the instanton-induced interactions
between fermions, and for $N_f$ flavours are the specific nonlocal
$2N_f$-fermion vertices
\beqa
Y^{\pm}_{N_f}[\psi,\psi^\dagger]\! &\pj=&\pj \!(-)^{N_f}\!\!
\int\!\!d^4z\:dU\! \prod^{N_f}\!\int\!d^4x\:d^4y\nonumber\\
&& \cdot\psi^\dagger_{L,R}(x)(i\partial-i\mu)^{\mp}\Phi_{\bar I,I}(x-z)
\tilde\Phi_{\bar I,I}(y-z)(i\partial-i\mu)^{\pm}\psi_{L,R}(y) .
\eeqa
Here we use the notation $x^{\pm}=x^\mu\sigma_\mu^{\pm}$, where the $2
\times 2$ matrices $\sigma_\mu^{\pm} = (\pm i \vec \sigma,1)$ decompose
the Dirac matrices into chiral components,
and it is understood that $\mu$ written as a four-vector is
$\mu_\alpha = (\vec 0,\mu)$.  Note that the zero mode in the field of
an (anti) instanton couples to that of a (right-) left-handed quark, and
the effective range of the interaction is the average instanton size,
$\bar\rho$.

It should be stressed that correlations between
instantons induced by fermions are inherent in this approach; as to
correlations induced by gluons, they are effectively taken care of by
the use of the variational principle \cite{DP1,DPW} resulting in the
effective size distribution.
For simplicity we freeze all
the sizes at the average value $\bar\rho$, but average explicitly over random
position ($z$) and orientation ($U$) variables.

Fermion operators in the pre-exponent are not
convenient; these operators can be raised into the exponent
with the help of a supplementary integration over a pair of Lagrange
multipliers, denoted $\lambda_{\pm}$:
\begin{eqnarray}
{\cal Z} &\pj=&\pj \int\!d\lambda_+d\lambda_-\int
\!D\psi\:D\psi^\dagger\:\exp\Bigg\{\int\!d^4x\:\psi^\dagger(i\dd-i\mu\gamma_4)
\psi + \lambda_+ Y^+_{N_f} + \lambda_- Y^-_{N_f}\nonumber\\
&& \quad\quad\quad\quad + N_+\left(\ln\frac{N_+}{\lambda_+ V} -1 \right) +
 N_-\left(\ln\frac{N_-}{\lambda_- V} -1 \right) \Bigg\}.
\label{Z5}
\end{eqnarray}
Indeed, integrating over $\lambda_\pm$ by the saddle-point
method one recovers \eq{Z4}.  Notice that the saddle-point
integration becomes exact
in the thermodynamic limit: $N_\pm,V \rightarrow\infty$ with $N/V$ fixed.
Through this procedure we obtain a purely exponential integrand which is
the required effective action.

Two important consequences follow and should be
emphasized.  First, the coupling constant
of the $2N_f$-fermion interaction, whose role is played by
$\lambda_\pm$, is not fixed once and forever; its value
is found from minimizing the free energy {\em after} integration over
fermions is performed. Therefore, the strength of the interaction
depends itself on the phase the fermion system assumes.
Second, the saddle-point value of $\lambda_\pm$ is {\it not}
proportional to the instanton density, as one might naively think.
For example, in the case of chiral symmetry breaking it behaves as
$\lambda\sim (N/V)^{1-N_f/2}$ \cite{DP4}.  Both circumstances are due to
the pecularity of the instanton-induced interactions dominated by the
existence of the zero modes. Would these modes remain zero, the fermion
determinant would be zero, thus suppressing the presence of instantons
themselves. It is this intricate fermion self-supporting mechanism
which makes the instanton-induced interactions different from a
more naive NJL model where the strength of the effective four-fermion
interactions are chosen once and forever. As will be shown below, this
is the mechanism by which chiral symmetry is restored at large chemical
potential in the case of two flavours.

For practical applications it is
favourable to use the Fourier-transformed expressions of the quark zero
modes, which are written explicitly in the Appendix.
 These complex
functions of the four-momenta and chemical potential determine the
(matrix) formfactors attached to each fermion leg of the vertex:
\beq
{\cal F}(p,\mu) = (p+i\mu)^- \varphi(p,\mu)^+\, , \quad\;\;\;
{\cal F}^\dagger(p,-\mu) = \varphi^*(p,-\mu)^-(p+i\mu)^+\,.
\la{calF}\eeq
With such definitions, the interaction terms may be written in momentum space,
\beqa
Y^+[\psi,\psi^\dagger]  &\pj=&\pj \int\!dU\;
\int\!\prod_f^{N_f}\left[\frac{(d^4p_fd^4k_f)}{(2\pi)^{8}}\right]
(2\pi)^4\delta^4\Big(\sum(p_f-k_f)\Big) \nonumber\\
&&\hspace{-1.4cm}\cdot \prod^{N_f}_f \Big[\psi^\dagger_{Lf\alpha_f i_f}(p_f)
{\cal F}(p_f,\mu)_{k_f}^{i_f}\epsilon^{k_fl_f} U_{l_f}^{\alpha_f}
 U_{\beta_f}^{\dagger o_f} \epsilon_{n_fo_f}
 {\cal F}^\dagger(k_f,-\mu)_{p_f}^{n_f}\psi_L^{f\beta_fp_f}(k_f)\Big]\,
\la{genvertex}\eeqa
with a similar form for $Y^-$ which carries right-handed quarks.  The first
indices on the fermion operators refer to flavour, the Greek to colour
($1\dots N_c$), and the last denote spin ($1,2$).  This formulation of the
effective interaction retains the full $p$ and $\mu$ dependence of the zero
modes, as opposed to the approximate treatments in other recent works
\cite{StonyBrook,Princeton/MIT}.  At the same time this formalism
enables one to reduce the problem of determining the phase structure
to algebraic equations where the coefficients are given by certain integrals
over the formfactor functions \ur{calF}.

\section{One Light Flavour (A Pedagogical Aside)}

In the case of one flavour, the effective action assumes a relatively
simple and hence instructive form.  The formfactors combine and appear
as an effective mass term.  The interaction remains non-local, thus
one finds an induced mass, ${\cal M}$, with a specific momentum dependence:
\beqa
S_{INT} &\pj=&\pj -\lambda_+\int\!\frac{d^4p}{(2\pi)^4}\int\! dU \:
\psi^\dagger_{L\alpha i}(p)
{\cal F}(p,\mu)_{j}^{i}\epsilon^{jk} U_{k}^{\alpha}
 U_{\beta}^{\dagger l} \epsilon_{ml}
 {\cal F}^\dagger(p,-\mu)_{n}^{m}\psi_L^{\beta n}(p)\nonumber\\
&& - \lambda_- (L \leftrightarrow R) \nonumber\\
&\pj= &\pj  {\cal M}(p,\mu)\psi^\dagger(p) \psi(p) \,,
\eeqa
where
\begin{equation}
{\cal M}(p,\mu) = \lambda (p+i\mu)_\alpha(p+i\mu)_\alpha \varphi_\beta(p,\mu)
\varphi_\beta(p,\mu) \,.
\end{equation}
At finite $\mu$ the formfactors are complex, since the zero modes are
solutions of \eq{diraceq}.
This interaction is sufficient to spontaenously break chiral symmetry.
However, the coupling constant $\lambda$ has here been introduced as a
Lagrange multiplier in order
to obtain a manageable form for the effective action,
and is in principle an integration over all possible coupling strengths.
This simplifies to a saddle-point approximation
which becomes exact in the thermodynamic limit of $N,V\rightarrow
\infty$.  Such an evalution of the integral also naturally connects the
instanton density to this coupling, in that the resulting `gap' equation
is:
\beq
\frac{N}{V} = 4 N_c \int \dbar{p} \frac{{\cal M}(p,\mu)^2}
{(p+i\mu)^2 + {\cal M}(p,\mu)^2 } \,.
\label{nf1gap}
\eeq
This equation is a direct generalization of the gap equation
found for $\mu=0$ in ref. \cite{DP3}.
Although ${\cal M}$ and the $\mu$ term are complex, one finds that
the imaginary part of the integrand is odd in $p_4$ and hence vanishes
under energy integration.  This is equivalent to the statement that while
individual eigenvalues of the Dirac operator are imaginary, their product
remains real.  By preserving the exact form of the zero mode solution,
this property becomes manifest in our treatment.
Through solving this equation self-consistently, one obtains a solution
for the $\lambda$ at any given quark chemical potential $\mu$.  This in
turn is proportional to an effective mass evaluated in the zero momentum
limit, $|{\cal M}(0,\mu)| = (2\pi\bar\rho)^2\lambda/N_c$.

To first order in $\lambda$, this procedure amounts to averaging the
effects of every instanton and anti-instanton which modify the
quark propagator.  In principle, the instanton density $N/V$ is modified by
a finite quark density.  It scales with the fugacity, $\chi \equiv
Z(\mu)Z(0)^{-1}$, as $\chi^{4/b}$, where $b$ is the Gell-Mann-Low
coefficient of QCD.  Fugacity not equal to unity simply arises
from the change with $\mu$ of the fermion determinant in the
partition function.  Upon a systematic expansion of this determinant in
$\lambda$ one finds the first order term vanishes, for this is exactly the
saddle-point condition of \eq{nf1gap}.
Thus modifications of the instanton background appear only when the
determinant is expanded to order $\lambda^2$, which is beyond the scope of
this work.

Without a first-order effect on $N/V$, there exists no mechanism in the theory
to surpress the instanton background and modify the form of the gap equation
above.  While a finite quark chemical potential can and does modify the
coupling strength and hence the effective mass, the coupling is never forced
to zero.  Therefore chiral symmetry remains broken at any finite baryon
density.  This result can of course change when effects of higher order in the
instanton density are taken into account.  The picture is also radically
altered with more than one flavour, as possible quark pairing offers
an alternative mechanism for chiral symmetry restoration.

\section{Two Light Flavours}

The remainder of this paper investigates the
case of massless $N_f = 2$, corresponding to a system of
chiral up and down quarks.  Furthermore, we will assume the CP
invariant case of $\theta=0$, which requires $N_+ = N_- = N/2$ and
hence $\lambda_+=\lambda_-=\lambda$.
With the definitions and notation of \eq{calF} and \eq{genvertex},
the interaction terms may be written in momentum space,
\beqa
\lambda Y^+ &\pj=&\pj \lambda
\int\!\frac{d^4p_1d^4p_2d^4k_1d^4k_2}{(2\pi)^{16}}
(2\pi)^4\delta^4(p_1+p_2-k_1-k_2) \nonumber\\
&&\cdot \int\!dU\:
\psi^\dagger_{L1\alpha_1 i_1}(p_1)
{\cal F}(p_1,\mu)_{k_1}^{i_1}\epsilon^{k_1l_1} U_{l_1}^{\alpha_1}
 U_{\beta_1}^{\dagger o_1} \epsilon_{n_1o_1}
 {\cal F}^\dagger(k_1,-\mu)_{p_1}^{n_1}\psi_L^{1\beta_1p_1}(k_1)
\nonumber \\
&&\quad\cdot \psi^\dagger_{L2\alpha_2 i_2}(p_2)
{\cal F}(p_2,\mu)_{k_2}^{i_2}\epsilon^{k_2l_2} U_{l_2}^{\alpha_2}
 U_{\beta_2}^{\dagger o_2} \epsilon_{n_2o_2}
 {\cal F}^\dagger(k_2,-\mu)_{p_2}^{n_2}\psi_L^{2\beta_2p_2}(k_2)\,,
\la{vertex}\eeqa
with a similar form for $\lambda Y^-$.

\subsection{Gap Equations}

Since the instanton-induced interactions \ur{vertex} support both
$\bar q q $ and $q q $ condensation, it is necessary to consider the two
competing channels simultaneously. This means that one must calculate
both the normal ($S$) and anomalous ($F$) quark Green functions. A
colour/flavour/spin ansatz compatible with the possibility of chiral
and colour symmetry breaking is
\beqa
\langle\psi^{f\alpha i}(p)\psi^\dagger_{g\beta j}(p)\rangle &\pj=&\pj
\delta^f_g \delta^\alpha_\beta S_1(p)^i_j \quad
{\rm for}\:\alpha,\beta=1,2 \,,
\nonumber
\\
\langle\psi^{f\alpha i}(p)\psi^\dagger_{g\beta j}(p)\rangle &\pj=&\pj
\delta^f_g \delta^\alpha_\beta S_2(p)^i_j \quad
{\rm for}\:\alpha,\beta>2 \,,
\nonumber
\\
\langle\psi_L^{f\alpha i}(p)\psi_L^{g\beta j}(-p)\rangle
&\pj=&\pj
\langle\psi_R^{f\alpha i}(p)\psi_R^{g\beta j}(-p)\rangle
= \epsilon^{fg}\epsilon^{\alpha\beta[\gamma]}\epsilon^{ij} F(p)\,,
\eeqa
where $[\gamma]$ refers to some generalized direction(s) in colour space,
and it is this set of $N_c-2$ indices which signals the breaking of
colour symmetry.  In the particular case of $N_c=3$, where the colour
symmetry is broken as $SU(3)\rightarrow SU(2)\times U(1)$ and our
ansatz considers the $\bar{3}$ channel, we will by
convention take $[\gamma]=3$; for $N_c=4$ one can take
$[\gamma]=34$ and so forth. In the event of colour symmetry
breaking, the standard propagators (and ensuing condensates) will lose
their colour degeneracy and the separation of $S(p)$ into $S_1(p)$ and
$S_2(p)$ becomes necessary; otherwise the Schwinger-Dyson-Gorkov equations do
not close.

Written in the chiral $L,R$ basis, the $4\times 4$ propagators
$S_{1,2}(p)$ are of the form:
\beq
S(p) = \left[\begin{array}{cc}G(p) {\bf 1} & Z(p){\bf S}_0(p)^+ \\
Z(p){\bf S}_0(p)^- & G(p) {\bf 1} \end{array}\right] \,.
\eeq
Here the off-diagonal, bare propagator
${\bf S}_0(p)^{\pm} = \left[(p+i\mu)^\pm\right]^{-1}$
is modified by the scalar functions $Z_{1,2}(p)$, and is augmented on the
diagonal by the scalar $G_{1,2}(p)$ which if nonzero break chiral symmetry.

\begin{figure}[bt]
\setlength\epsfxsize{13cm}
\centerline{\epsfbox{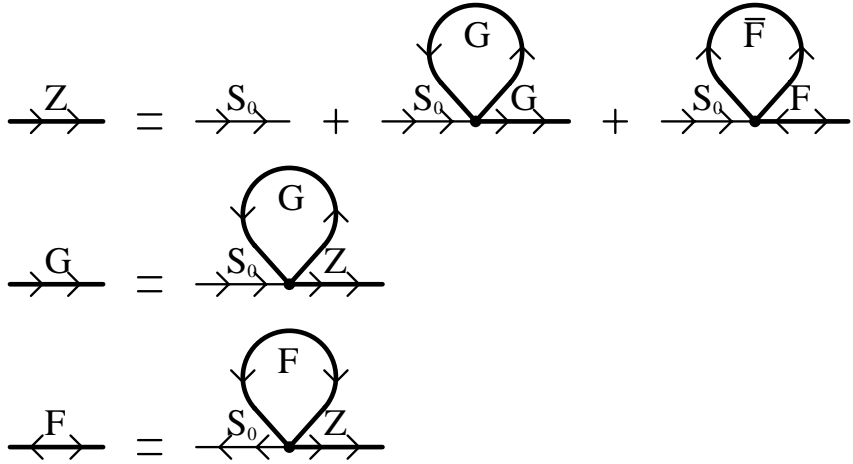}}
\caption{Schwinger-Dyson-Gorkov diagrams to first order in $\lambda$,
corresponding to five scalar equations in the text.  The first is for
the off-diagonal and the second for the diagonal components of $S_1$ and
$S_2$, the third is the anomalous diquark propagator.}
\label{gorkfig}
\end{figure}

Using the instanton-induced interaction \ur{vertex} one can build a
systematic expansion for the $F,G$ Green functions in the $1/N_c$
and $\bar\rho/\bar R$ parameters. In the leading order in both
parameters we restrict ourselves to the one-loop approximation
shown in \fig{gorkfig}. It corresponds to a set of self-consistent
Schwinger-Dyson-Gorkov equations.
An important $\mu$-dependence enters through the formfactors in
\ur{vertex}.
With $\bar F(p)=F^*(p)$ these diagrams lead to the set of five
algebraic equations for the scalar Green functions $Z_{1,2},\,G_{1,2}$
and $F$:
\beqa Z_1(p) &\pj=&\pj 1 - G_1(p) A(p,\mu)M_1 -
2 F(p)B(p,\mu)\Delta\nonumber\\ Z_2(p) &\pj=&\pj 1 - G_2(p)
A(p,\mu)M_2 \nonumber\\ G_1(p) &\pj=&\pj Z_1(p)
\varphi_\alpha(p,\mu)\varphi_\alpha(p,\mu)M_1 \nonumber \\ G_2(p)
&\pj=&\pj Z_2(p) \varphi_\alpha(p,\mu)\varphi_\alpha(p,\mu)M_2
\nonumber\\
F(p) &\pj=&\pj 2 Z_1(-p) \varphi_\alpha(p,\mu)\varphi_\alpha(-p,\mu)\Delta\,.
\la{five}\eeqa
The constants $M_1$, $M_2$, and $\Delta$ will be
defined below and the functions
\beqa
A(p,\mu) &\pj=&\pj (p+i\mu)_\alpha(p+i\mu)_\alpha \varphi_\beta(p,\mu)
\varphi_\beta(p,\mu) \,, \nonumber\\
B(p,\mu) &\pj=&\pj (p^2+\mu^2)\varphi_\beta(p,\mu)\varphi_\beta(-p,\mu)
+ (p+i\mu)_\alpha\varphi_\alpha(p,\mu)(p-i\mu)_\beta\varphi_\beta(-p,\mu)
\nonumber\\
&&\, - (p+i\mu)_\alpha\varphi_\alpha(-p,\mu)(p-i\mu)_\beta\varphi_\beta(p,\mu)
\,, \label{GFs}
\eeqa
are the formfactors which arise from the zero modes (see the
Appendix). At $\mu =0$ we have $A(p,0)=B(p,0)$,
but for any finite $\mu$ the direction of the momentum flow
through each vertex leg is critical.

The condensates $g_1$, $g_2$, and $f$ are the closed loops
contributing to the quark self-energy.  They are found by integrating the
appropriate Green function, modified by the vertex formfactors, over
an independent momentum:
\beqa
g_{1,2} &\pj=&\pj \frac{\lambda}{N_c^2-1} \int\!\frac{d^4k}{(2\pi)^4}\:
A(k,\mu)G_{1,2}(k) \,,\nonumber\\
f &\pj=&\pj \frac{\lambda}{N_c^2-1}
\int\!\frac{d^4k}{(2\pi)^4}\: B(k,\mu)F(k) \,.
\label{condefs}
\eeqa
The quantities $M_{1,2}$ and $\Delta$ are linear
combinations of the condensates $g_{1,2}$ and $f$:
\beqa
M_1 &\pj=&\pj \left(5-\frac{4}{N_c}\right)g_1+
\left(2 N_c-5+\frac{2}{N_c}\right)g_2
\, , \nonumber\\
M_2 &\pj=&\pj 2\left(2-\frac{1}{N_c}\right)g_1+2(N_c-2)g_2\,,\nonumber\\
\Delta &\pj=&\pj \left(1+\frac{1}{N_c}\right)f\,.
\label{defs}
\eeqa
The $M_{1,2}$ are measures of chiral symmetry breaking, which
act as an effective mass modifying the standard quark propagation.  On the
other hand the diquark loop $2\Delta$ plays the role of twice
the single-quark energy gap formed around the Fermi surface.  The Fermi
momentum, in the absence of chiral symmetry breaking, will remain at
$p_f=\mu$ regardless of the magnitude of $\Delta$.
It should be noted that the effective masses $M_{1,2}$ exhibit
different behaviour in $N_c$ than the superconducting gap $\Delta$:
the former are $N_c$ times larger than the latter.  Although one can
of course express a solution in terms of $(g_1,g_2,f)$ as one in
$(M_1,M_2,\Delta)$, we will retain both notations in the
following text.  The reason is the distinct physical interpretations for
the two variable sets:  the first quantifies the symmetry breaking by
channel while the second measures its effects on the quark states with
particular colour.

After determining the five scalar functions through solving eqs. \ur{five}
and inserting these solutions into eqs. \ur{condefs} we find the coupled
equations for the condensates themselves:
\beqa
g_1 &\pj=&\pj \frac{\lambda M_1}{N_c^2-1}\int\!\frac{d^4k}{(2\pi)^4}\:
\frac{\alpha(p,\mu)
\left[1-4\beta(p,\mu)\Delta^2+\alpha^*(p,\mu)M_1^2\right]}
{\left[1+\alpha(p,\mu)M_1^2\right]
\left[1+\alpha^*(p,\mu)M_1^2\right]-16\beta(p,\mu)^2\Delta^4}\,, \nonumber\\
g_2 &\pj=&\pj \frac{\lambda M_2}{N_c^2-1}\int\!\frac{d^4k}{(2\pi)^4}\:
\frac{\alpha(p,\mu)}{1+\alpha(p,\mu)M_2^2}\,,\nonumber\\
f &\pj=&\pj \frac{2\lambda\Delta}{N_c^2-1}\int\!\frac{d^4k}{(2\pi)^4}\:
\frac{\beta(p,\mu)\left[1-4\beta(p,\mu)\Delta^2+\alpha(p,\mu)M_1^2\right]}
{\left[1+\alpha(p,\mu)M_1^2\right]
\left[1+\alpha^*(p,\mu)M_1^2\right]-16 \beta(p,\mu)^2\Delta^4}\,,
\label{CONDS}
\eeqa
where yet another pair of functions has been introduced,
\beq
\alpha(p,\mu) = A(p,\mu)\varphi_\alpha(p,\mu)\varphi_\alpha(p,\mu)\,,\quad
\beta(p,\mu) = B(p,\mu)\varphi_\alpha(p,\mu)\varphi_\alpha(-p,\mu)\,.
\eeq
Note that while $\beta$ is real, the function $\alpha$ is complex.
Although the integrands above are complex, all imaginary parts are
odd in $p_4$ and thus vanish under integration.  As usual
for gap equations, there is the possibility of a solution where some (or
even all) of the condensates vanish.

\begin{figure}[tb]
\setlength\epsfxsize{10cm}
\centerline{\epsfbox{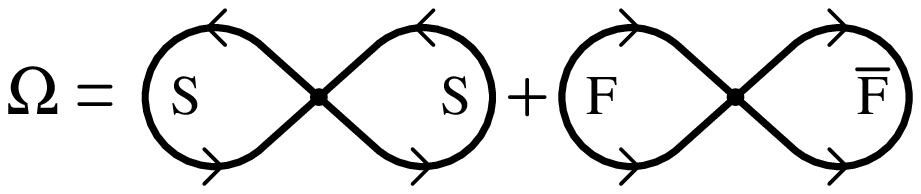}}
\caption{Contributing diagrams to $\Omega$ at order $\lambda^1$.}
\label{eightfig}
\end{figure}

The magnitudes of these condensates are not yet determined, as this
requires fixing the coupling constant $\lambda$.  This is done through
minimizing the partition function \ur{Z5} and leads to the saddle point
condition on $\lambda$, replacing the integration with
\beq
\frac{N}{V} = \lambda\langle Y^+ + Y^- \rangle \,.
\label{pregap}
\eeq
On the left-hand side is the instanton density, which we will here take
to be fixed at its vacuum value, although in principle it will have some
correction due to the finite quark density.  Evaluating the right-hand
side requires calculating the one-vertex contributions to the free
energy, which in this case includes two-loop, figure-eight type diagrams
formed by joining the four fermion legs into two pairs as
shown in \fig{eightfig}.  This can be
written concisely in terms of the condensates:
\beq
\frac{N}{V} = \lambda\langle Y^+ + Y^-\rangle = \frac{4(N_c^2-1)}{\lambda}
\left[ 2 g_1 M_1 + (N_c-2) g_2M_2 + 4 f\Delta \right]\,.
\label{gapeqn}
\eeq
As with the single flavour case we are quenching by
truncating our perturbative treatment in $\lambda$, and thus the instanton
denisty on the left is held fixed.
This equation for $\lambda$ and the definitions of the condensates
\ur{CONDS} comprise a system of equations which can be solved for all
quantities.
It follows from eqs. \ur{defs}, \ur{CONDS} and \ur{gapeqn} that the mass and
energy gaps $M_{1,2}$ and $\Delta$ are proportional to the
{\it square root} of the instanton density, $N/V$, while the
saddle-point value of $\lambda$ is, to the first approximation,
independent of $N/V$.

Once these quantities are found the chiral condensate may be calculated,
being a trace over the quark propagator.  This is distinct from the
effective masses $M_{1,2}$, as it does not contain contributions from the
formfactors present in the loops from which the effective masses are
computed.  Present instead is the solution for the
propagator, in particular its diagonal elements:
\beqa
 -\langle\bar{\psi}\psi\rangle_{Mink} =
i\langle\psi^\dagger\psi \rangle_{Eucl} &\pj=&\pj
i\int\!\frac{d^4p}{(2\pi)^4}\,\Tr{S(p)} \nonumber \\
&\pj=&\pj 4\int\!
\frac{d^4p}{(2\pi)^4}\,\left[2 G_1(p) + (N_c-2) G_2(p)\right] .
\eeqa
As with the previous integrals, the integrand reflects in $p_4$ such that
the result is guaranteed real.

\subsection{Thermodynamic Competition}

As a consequence of having two different modes of quark condensation,
one obtains multiple solutions for the characteristics of the quark
medium at any fixed chemical potential.  Specifically, there will be a
competition between the following phases:

{\it (0)  Free massless quarks:} $g_1=g_2=f=0$.

{\it (1)  Pure chiral symmetry breaking:} $g_1=g_2\ne 0$, $f=0$.  This is the
standard vacuum scenario.

{\it (2)  Pure diquark condensation:} $g_1=g_2=0$, $f\ne 0$.  This is the
`colour superconducting' phase, and leaves quarks of colour $[\gamma]$
free, since they neither participate in the diquark formation nor exhibit
broken chiral symmetry.

{\it (3)  Mixed symmetry breaking:}  $g_1 \ne g_2 \ne 0$, $f\ne 0$.  In fact
one need not have finite $g_1$, since it corresponds to chiral symmetry
breaking by the quarks of transverse colour to the diquark condensate,
however such a phase requires that there at least be chiral breaking in
the parallel-coloured quarks.

Phase (0), with all symmetries restored, would mean that the average
four-fermion vertex, $\langle Y^+ +Y^-\rangle$, is zero. It would imply,
via eq. \ur{Z5}, that the saddle-point value of $\lambda$ is infinite. Being
substituted into eq. \ur{pregap}, infinite $\lambda$ means that the fermion
determinant vanishes. Consequently, the whole instanton vacuum is
severely supressed, which means a large loss in the free energy. One
can put it in another way: if the density of instantons $N/V$ is
considered fixed, there is no solution to the combined eqs.
\urss{defs}{CONDS}{pregap}
with all condensates being zero, at least in the one-loop approximation
to the Schwinger-Dyson-Gorkov equation we are considering.

To resolve between the remaining, symmetry-breaking solutions
the free energy is minimized. Consistent with the
evaluation of the Green functions, it is calculated to first order in
$\lambda$. Repeating the calculation of figure-eight diagrams and
recalling the explicit dependence on $\lambda$ in \eq{Z5} we obtain the
free energy
\beqa
\frac{\Omega}{V_3} &\pj=&\pj -\frac{1}{\beta V_3}\ln Z \nonumber\\
&\pj=&\pj \frac{\Omega_0}{V_3}-\frac{N}{V}\ln\left(\frac{N}{\lambda V}\right)
+ \frac{N}{V}
-\frac{4(N_c^2-1)}{\lambda}
\left[ 2 g_1 M_1 + (N_c-2) g_2M_2 + 4 f\Delta \right].
\eeqa
Here $V_3$ is the three-volume while $V$ is the Euclidean four-volume, and
$\Omega_0$ is the free energy for a gas of free quarks.
The last two terms are precisely the quantity which must vanish under the
saddle-point determination of $\lambda$, and thus we have
\beq
\frac{\Omega}{V_3} = \frac{\Omega_0}{V_3}+\frac{N}{V}
\ln\left(\frac{\lambda}{N/V}\right)\,.
\eeq
Thus the phase which features the {\em lowest} coupling $\lambda$
is the thermodynamically favoured.

If we first concentrate on a competition between the two phases of simple
chiral or colour symmetry breaking,
(1) and (2), the former is marked by nonzero $g\equiv g_1=g_2$ and
the latter by nonzero $f$. Each case leads to its own value of
$\lambda$ through solving \eq{gapeqn} for any given $\mu$.
For the first, the gap equation reduces to
\beq
\lambda = \frac{8(N_c^2-1)^2 g^2}{N/V} \,,
\eeq
while for the second one finds
\beq
\lambda = \frac{16 (N_c^2-1)(N_c+1) f^2}{N_c (N/V)}\,.
\eeq
At certain value of $\mu=\mu_c$ these solutions for $\lambda$
cross; this is the point where
the phase transition occurs.  This point is defined by the condition
\beq \frac{f(\mu_c)}{g(\mu_c)} = \sqrt{\frac{N_c(N_c-1)}{2}}\,.
\label{CR} \eeq
When the ratio between the condensates is less than the constant on the
right, phase (2) is favoured; otherwise it is chiral symmetry that is
spontaneously broken in phase (1).
However, the phase structure is not so simple in general, considering the
possible presence of phase (3).
Calculations were carried out for
$N_c = 2$ and $N_c = 3$, taking the values for $N/V$ and $\bar\rho$ specified
in Section 2.  From eq. \ur{CR} one observes that at large $N_c$ it becomes 
increasingly difficult to get to the colour superconducting phase.

\section{Two Colours}

In this case it is obvious that colour symmetry is not broken by diquark
formation, which here correspond to colour-singlet `baryons', and hence
there is only one possible chiral condensate $g_1=g$.
Not so obvious is that at $\mu =0$ the colour and flavour
$SU(2)$ groups are arranged into the higher $SU(4)$ symmetry.  So
no mixed phases exist; the theory has only one symmetry.
The instanton vacuum accounts for this \cite{DP5}, and in
the context of the analysis here this corresponds to
$f^2 + g^2$ being the only discernable quantity in the gap
equations \ur{CONDS}. Numerically, we find $\sqrt{f^2+g^2}= 147$ MeV.

At finite $\mu$, however, the $SU(4)$ symmetry is explicity broken and
there is a thermodynamic competition between phases of pure chiral or
diquark condensation.  For all finite chemical potential, numerical
calculations reveal that $f$ decreases faster than $g$.  Since in this case the
critical ratio of \eq{CR} is unity, we conclude that for any finite
density the $N_c=2$ world prefers diquark condensation to chiral
symmetry breaking.  This finding is in agreement with the reasoning of
ref. \cite{StonyBrook} and the lattice results of ref. \cite{Nc2Lattice}.

\section{Three Colours}

With three colours, the more relevant case for QCD,
the phase structure is richer due to the possible
appearance of phase (3).  Furthermore, the physics is naturally completely
different, since here any diquark formation will spontaneously prefer a
particular colour.  This dynamical consequence mimics the Higgs mechanism
as a spontaneous breaking of the $SU(N_c)$ gauge group.

\subsection{Numerical Results}

The system of gap equations is comprised of all three
in \eq{CONDS}, which are accompanied by the condition (\ref{gapeqn}).
For a given $\mu$ a set of solutions can be obtained numerically,
each of which corresponds to a phase of the quark matter characterized by
the mass or energy gaps formed.

In the vacuum, where $\mu$ strictly vanishes, there exist solutions for all
three symmetry-breaking phases.  The thermodynamically favoured is one of
spontaneous chiral symmetry breaking, and hence the equilibrium description
of the vacuum in the instanton picture is recovered. Numerical
calculations show that the constituent quark mass is $M\equiv M_1=M_2=346$ MeV,
corresponding to $\lambda^{(1)} = 0.311$.  Two metastable states
are also present, these being manifestations of phases (2) and (3).
They feature $\lambda^{(2)} = 0.674$ and $\lambda^{(3)} = 0.673$, and are
thus clearly only local minima of the free energy.  It is noteworthy that
the diquark gap in the metastable state is quite large: $\Delta = 220$ MeV.
This is in agreement with that found for $\mu=0$ in 
ref.\cite{DFL}\footnote{In that work the notation $m_c=2\Delta$ was used.}.

\begin{figure}[tb]
\setlength\epsfxsize{10cm}
\centerline{\epsfbox{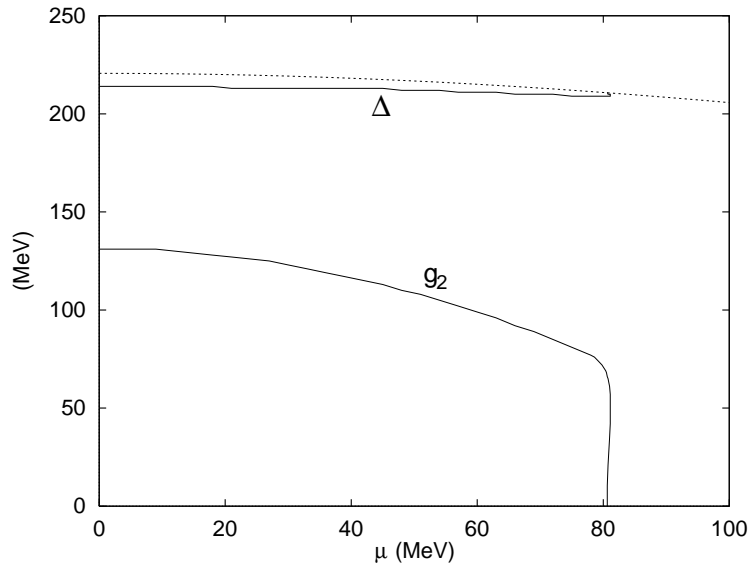}}
\caption{ Chiral and diquark gaps as a function of chemical potential
in the phase of mixed symmetry breaking, (3).
The dashed line is the phase (2) solution of the diquark energy gap.}
\label{ph3}
\end{figure}

Although it is an unstable state in vacuum and remains so as $\mu$ is
increased, the phase of mixed symmetry breaking is worthy of attention.
It is marked by nonzero condensates $g_2$ and $f$, which leads to
not only a superconductivity energy gap but also finite and distinct
effective quark masses, $M_1$ and $M_2$.  The magnitudes of these gaps are
plotted as a function of chemical potential in \fig{ph3}.
In effect here is a separation between the phenomenology of quarks of
different colours.  Quarks orthogonal in colour to the diquark condensate
(colours 1 and 2)
do not conspire to break chiral symmetry (so $g_1=0$), and rather condense in
colour-asymmetric pairs.   This leaves the quarks colour-parallel (colour
3) to the diquark $\bar{3}$ to spontaneously break chiral symmetry in the
usual manner ($g_2\ne 0)$.  Thus they acquire an effetive mass due to
self-interaction and bestow a mass upon the other two quarks through a similiar
interaction.  Since the SU(3) colour symmetry has been broken, these masses
need not be equivalent, and in fact one finds that
\beq
M_1 = \frac{5}{3}g_2 \,,\qquad M_2 = 2 g_2 \,.
\eeq
In the vacuum, $M_1 = 220$ MeV and $M_2 = 260$ MeV.
As the chemical potential increases, the coexisting diquark and chiral
condensates compete for the coupling strength; this is the
condition of \eq{gapeqn}.
We find that at relatively low $\mu$, around
80 MeV, this limited resource becomes entirely consumed by the diquark.  This
is demonstrated in \fig{ph3}, where one notes the chiral condensate in the
colour-3 channel abruptly vanishes and the diquark condensate assumes
its value of phase (2).  Thus this solution is not only strictly a
local minimum, but is also short-lived in density.

The remaining thermodynamic competition is thus between the two phase of
pure chiral or colour symmetry breaking.
For three colours, the critical ratio of the condensates
\ur{CR} is $\sqrt{3}$. At $\mu =0$ we find
$f/g=165\,{\rm MeV}/65\,{\rm MeV} > \sqrt{3}$. It means that at
low values of $\mu$ the coupling constant $\lambda$ is smaller in
the chiral broken phase, hence this phase is energetically preferred.
Furthermore, this is the $\lambda$ which leads to a reasonable
effective quark mass of $M = 346$ MeV, as well as a chiral condensate of
$\langle\bar\psi\psi\rangle = -(255\,{\rm MeV})^3$.  We thus recover the
standard chiral symmetry breaking at low baryon density.

\begin{figure}[tb]
\setlength\epsfxsize{10cm}
\centerline{\epsfbox{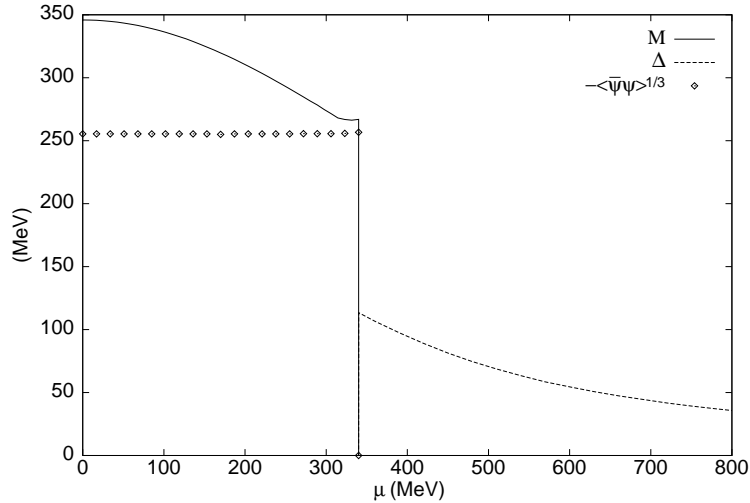}}
\caption{Condensates for $N_c=3$ as a function of $\mu$.
Shown are the effective quark mass $M$, the quark condensate
$-\langle\bar{\psi}\psi\rangle^{1/3}$, and the diquark energy gap
per quark $\Delta$.} \label{plot} \end{figure}

With increasing chemical potential both $f$ and $g$ condensates in their
respective phases are reduced, however as $\mu$ surpasses the effective
quark mass $M$ around 300 MeV, the chiral-breaking $g$ stabilizes
while the colour-breaking $f$ continues to decrease in phase (2).  At
$\mu\simeq 340$ MeV condition \ur{CR} is reached, and therefore for this
and greater chemical potential chiral symmetry is restored and the
`colour superconducting' phase is realized.  At the transition point,
the quark gap is $\Delta = 115$ MeV and slowly decreasing with
rising density.  Numerical results for the characteristic quantities of
each phase are plotted in \fig{plot}.  It is noteworthy that the chiral
condensate, $\langle\bar{\psi}\psi\rangle$, is distinct from the
effective quark mass, $M$, in that while the latter decreases
with $\mu$ the former remains practically at its vacuum value. For
$\mu > \mu_c$ they both vanish.

\subsection{Properties of Quark Matter}

The chemical potential is a useful concept only in that it constrains the
particle number or density, the physical quantity we actually need to 
compute. The four-fermion interactions we are dealing with 
are somewhat unusual in that they are dependent on the energy, a 
consequence of using the zero mode Green functions as the seed of the 
interaction.

For this reason we carefully define the density as
\beq
n = \int\,d^4x\,j_4(x) \,,
\eeq
and note that in this case one does not have the simple form
$j_\mu=\psi^\dagger\gamma_\mu\psi$
for the current which carries conserved particle number.  In our notation
this corresponds to the current of the high-momentum quarks, which
without the zero modes would imply that the correct normalization is not
preserved.  We find the applicable current is
\beqa
j_{\mu}(q) &\pj=&\pj -i\psi^\dagger(q)\gamma_{\mu}\psi(q) + i\lambda
\int\!\frac{d^4p_1d^4p_2d^4k_1d^4k_2}{(2\pi)^{16}}
(2\pi)^4\delta^4(p_1+p_2-k_1-k_2) \nonumber\\
&&\cdot \int\!dU\: \psi^\dagger_{L1\alpha_1 i_1}(p_1)
\epsilon^{k_1l_1} U_{l_1}^{\alpha_1} U_{\beta_1}^{\dagger o_1}\epsilon_{n_1o_1}
\psi_L^{1\beta_1p_1}(k_1) \psi^\dagger_{L2\alpha_2 i_2}(p_2)
\epsilon^{k_2l_2} U_{l_2}^{\alpha_2}
U_{\beta_2}^{\dagger o_2} \nonumber\\ && \cdot
\epsilon_{n_2o_2} \psi_L^{2\beta_2p_2}(k_2)
\frac{\partial}{\partial q_{\mu}}\left[
{\cal F}(p_1,\mu)_{k_1}^{i_1}{\cal F}^\dagger(k_1,-\mu)_{p_1}^{n_1}
{\cal F}(p_2,\mu)_{k_2}^{i_2}{\cal F}^\dagger(k_2,-\mu)_{p_2}^{n_2}
\right] \nonumber\\
&&+ (L\leftrightarrow R)\,,
\eeqa
where the bracketed momentum derivative is understood to be a sum of
four terms, each with one formfactor differentiated with respect to its
momentum argument.

Although there has been some rearrangement, this is
essentially the vertex \ur{vertex} in which the formfactors have been
differentiated with respect to the momentum $q_{\mu}$.
One can verify this current is conserved by considering the Dirac equation,
written here for the left-handed, flavour-1 component:
\begin{eqnarray}
&&\hspace{-.8cm}\left[(p+i\mu)^-\right]^i_j\psi_L^{1\alpha j}(q) -
\lambda\int\frac{d^4p_2d^4k_1d^4k_2}{(2\pi)^{12}}
(2\pi)^4\delta(q+p_2-k_1-k_2)\nonumber\\
&&\cdot\int dU {\cal F}(q,\mu)_{k_1}^{i} U_{l_1}^{\alpha}
 U_{\beta_1}^{\dagger o_1} \epsilon_{n_1o_1}
 {\cal F}^\dagger(k_1,-\mu)_{p_1}^{n_1}\psi_L^{1\beta_1p_1}(k_1)\nonumber\\
&&\cdot \psi^\dagger_{L2\alpha_2 i_2}(p_2)
{\cal F}(p_2,\mu)_{k_2}^{i_2}\epsilon^{k_2l_2} U_{l_2}^{\alpha_2}
 U_{\beta_2}^{\dagger o_2} \epsilon_{n_2o_2}
 {\cal F}^\dagger(k_2,-\mu)_{p_2}^{n_2}\psi_L^{2\beta_2p_2}(k_2) = 0 \,.
\end{eqnarray}
Acting on this with $q_{\mu}\psi^\dagger_{L1}(q)\frac{\partial}
{\partial q_{\mu}}$, and with similar operations on similar equations for
the other components, one recovers the required $q_{\mu}j_{\mu}(q) = 0$.

For three colours we have found two phases which specify the stable
equilibria, each having one broken symmetry.
Phase (1) denotes spontaneously broken chiral symmetry, and it
competes with phase (2) which breaks colour symmetry.
As an intermediate step in calculating the density, we
first examine the quark occupation numbers for the two states, $n(p)$
where in this context $p \equiv |\vec{p}|$,
which is obtained via a $p_4$ integration and precisely defined as:
\beq
n = -i \int \frac{d^3p}{(2\pi)^3} \; n(p)\,.
\eeq
With this definition we have not separated the quark and anti-quark
distribution functions, since the phases we analyse do not necessarily retain
these as basic degrees of freedom.

The conserved current written above is a sum of two parts.  The first
is computed with the Green functions obtained through self-consistent
solution of the Schwinger-Dyson-Gorkov equations, and to the density
it contributes
\beqa
\int \dbar{p} \left(-\psi^\dagger(p)\gamma_4\psi(p)\right)
&\pj=&\pj -i \int \dbar{p} \Tr{S(p)\gamma_4} \nonumber\\
&\pj=&\pj -i \int \dbar{p} \left[\frac{p_4+i\mu}{(p+i\mu)^2}\right]
\left[2Z_1(p)+(N_c-2)Z_2(p)\right].
\eeqa
The scalar functions $Z_{1,2}(p)$ are complex, and for a given solution
set $(M_1,M_2,\Delta)$ become
\beqa
Z_1(p) &\pj=&\pj \frac{1+\alpha^*(p,\mu)M_1^2 - 4\beta(p,\mu)\Delta^2}
{\left[1+\alpha(p,\mu)M_1^2\right]\left[1+\alpha^*(p,\mu)M_2^2
\right] - 16\beta(p,\mu)^2\Delta^4} \,, \nonumber\\
Z_2(p) &\pj=&\pj \frac{1}{1+\alpha(p,\mu)M_2^2}\,.
\label{zsoln}
\eeqa
The functions $\alpha(p,\mu)$ and $\beta(p,\mu)$, as defined
previously, arise as combinations of the formfactors evaluated with
appropriate momentum arguments.  In general, the first is complex while
the second is purely real and always positive, a consequence of this being
the expression which accompanies the diquark propagation.  Through pairing
quarks, the complications due to the non-hermiticity of the single-quark
Dirac operator are avoided.

In the current's second term the formfactors have been differentiated
with respect to $p_4$.  The expectation values were computed
to the same ${\cal O}(\lambda^1)$ as followed for the Green functions, and
this generates another sibling $\gamma(p,\mu)$ in our family of functions.
\beq
\gamma(p,\mu) = (p+i\mu)_\alpha(p+i\mu)_\alpha \varphi_\beta(p,\mu)
\varphi_\beta(p,\mu) \varphi_\lambda \frac{\partial\varphi_\lambda(p,\mu)}
{\partial p_4} \,.
\eeq
Combining the two pieces, the density reads
\beqa
n &\pj=&\pj -iN_f \int \dbar{p} \Bigg\{ \left[\frac{p_4+i\mu}{(p+i\mu)^2}\right]
\left[2Z_1(p)+(N_c-2)Z_2(p)\right] \nonumber\\
&& - \frac{8}{\varphi_\alpha(p,\mu)\varphi_\alpha(p,\mu)}
\left[ \frac{p_4+i\mu}{(p+i\mu)^2} \alpha(p) + \gamma(p)\right]
\Biggl[\left(5-\frac{4}{N_c}\right)G_1(p)g_1 \nonumber\\
&& + \left(N_c-2\right)^2G_2(p)g_2 +
\left(2N_c-5+\frac{2}{N_c}\right)\left(G_1(p)g_2+G_2(p)g_1\right)\Biggr]
\Bigg\} \,.
\label{fulldens}
\eeqa
The functions $G_{1,2}(p)$ can be deduced from Eqs. \ur{five} and \ur{zsoln}.
The integrand here is complex, however under $p_4$ reflection we note that
$\alpha(-p_4) = \alpha^*(p_4)$ and $\gamma(-p_4)=-\gamma^*(p_4)$.
This follows from our preserving the non-hermitian contributions 
exactly, again we note that upon $p_4$ integrate we recover purely real 
occupation numbers and hence quark density.

\begin{figure}[tb]
\setlength\epsfxsize{10cm}
\centerline{\epsfbox{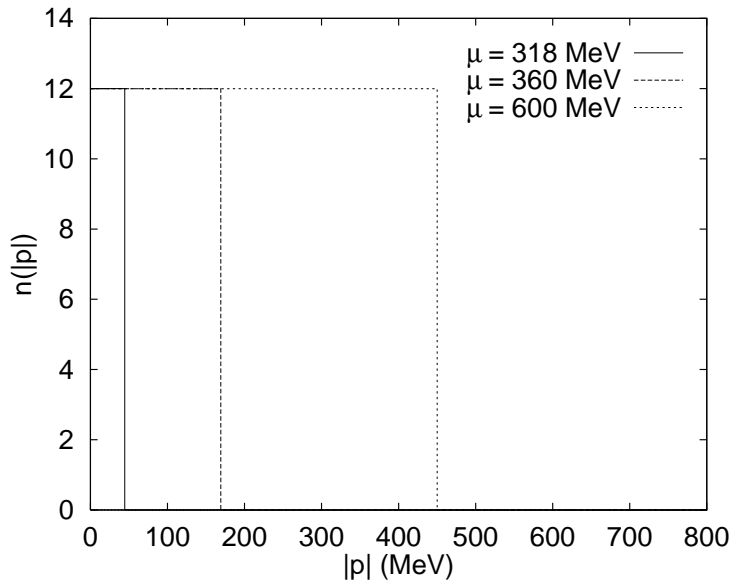}}
\caption{Occupation number $n(p)$ vs. $p$ in phase (1), where
$\mu =$ 318, 360, and 600 MeV.  The corresponding $p_F =$ 45, 170, and
450 MeV, respectively.}
\label{ph1dist}
\end{figure}

We now consider the two phases individually.  When phase (1) is assumed,
and $\Delta=0$, the density expression (\ref{fulldens}) simplifies to
a form where the integrand contains a denomenator of the form
$(p+i\mu)^2 + {\cal M}(p,\mu)^2$. A chiral effective mass is 
clearly present, which upon evaluation of \eqs{CONDS}{gapeqn} 
satisfies the same 
gap equation as in the $N_f=1$ case, \eq{nf1gap}\footnote{In fact the 
same gap equation applies to order $\lambda^1$ for any number of 
flavours \cite{DP4}.}.  The effective mass shifts the Fermi surface, 
which for free quarks is at $p_F=\mu$, to the solution of $p_F = 
\sqrt{\mu^2-{\cal M}(p_F,\mu)^2}$.  This, as shown in \fig{ph1dist}, is 
the only modification of the standard fermionic step function (of 
magnitude $2N_fN_c=12$) due to chiral symmetry breaking.

\begin{figure}[tb]
\begin{center}
\leavevmode
\setlength\epsfxsize{7.3cm}
\epsfbox{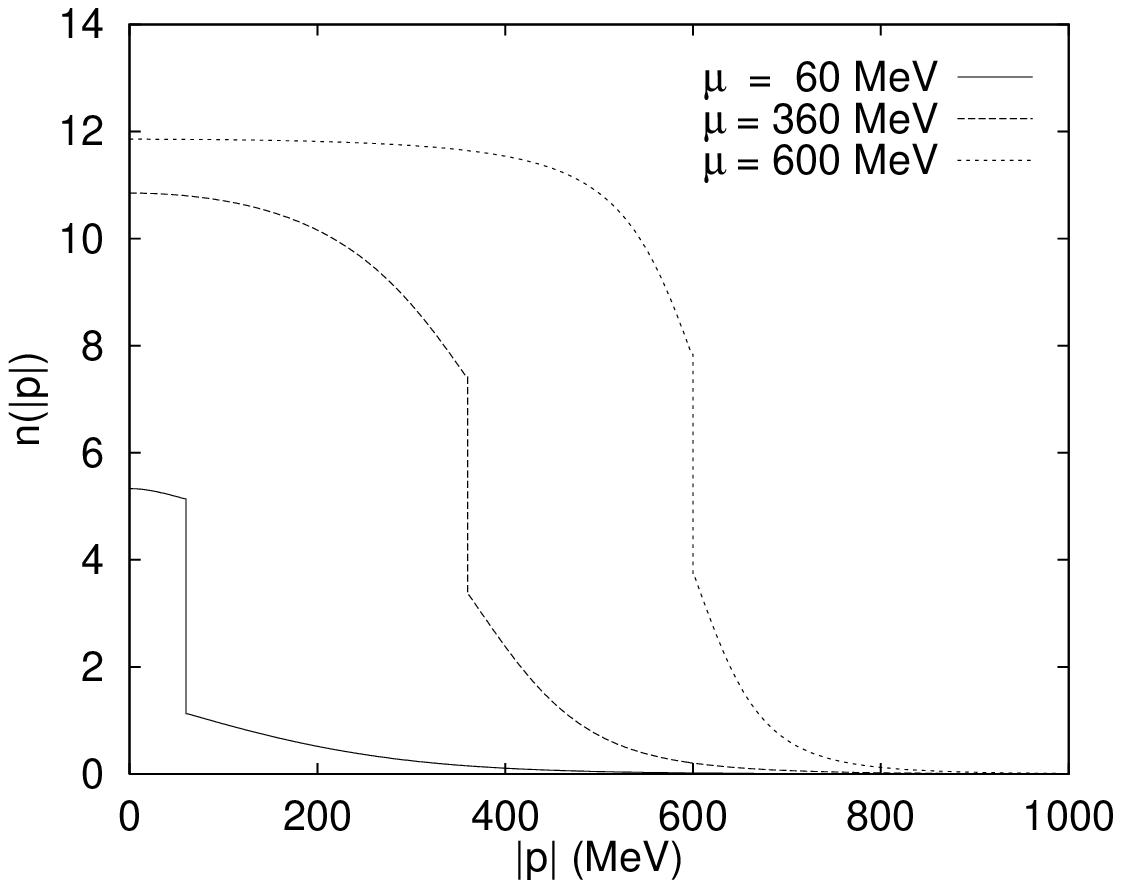}
\setlength\epsfxsize{7.3cm}
\epsfbox{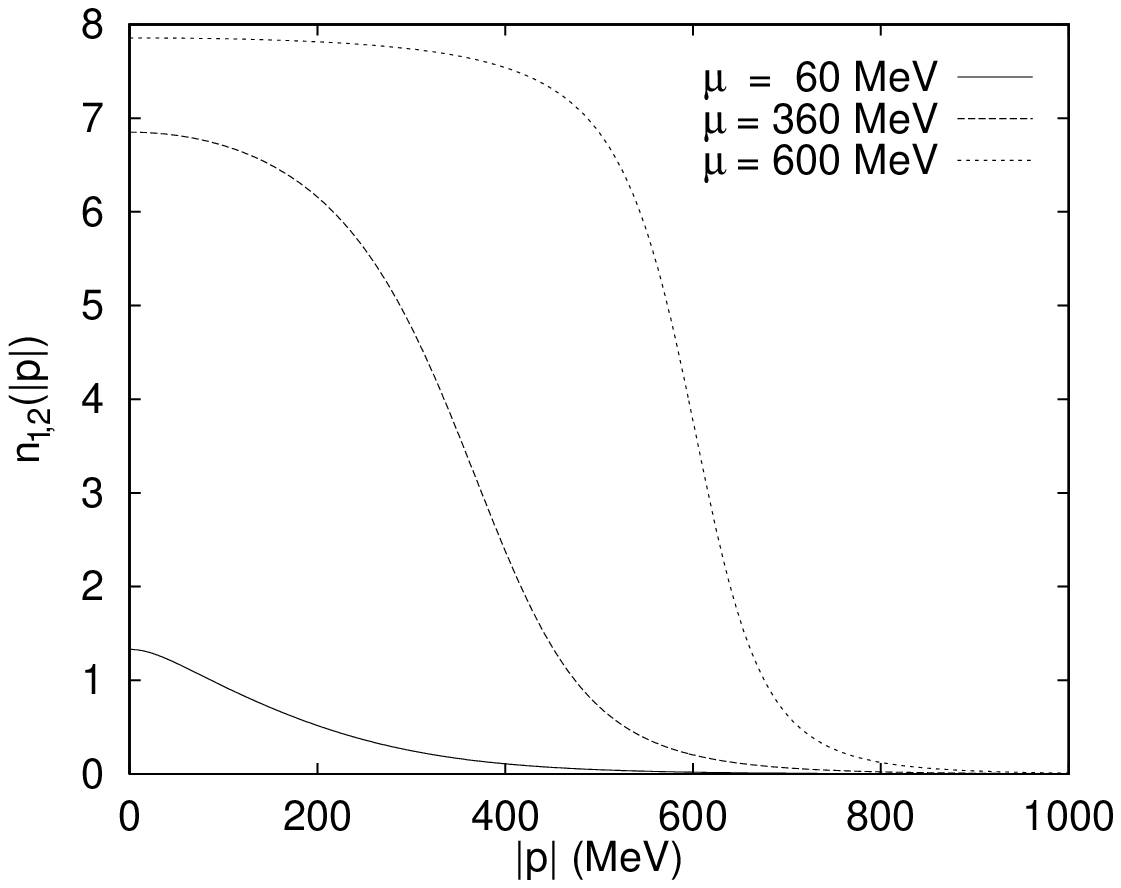}
\end{center}
\caption{Occupation number $n(p)$ vs. $p$ in phase (2), where
$\mu =$ 60, 360, and 600 MeV.  The plot on the left is for all three
colours together, the right for the two which participate in the bosonic
diquark condensate.}
\label{ph2dist}
\end{figure}

The occupation numbers are more interesting when the quarks fall into the
colour superconducting state.  As is known in BCS theory, Cooper pairing
of quarks spreads the Fermi surface as fermions pair with momenta above
and below the chemical potential.  As demonstrated in \fig{ph2dist}, this
is precisely the result for the two quark colours which participate in the
diquark.  Quarks of the parallel colour are unaffected by the energy gap,
and furthermore since chiral symmetry has been completely restored appear
as massless fermions with a degeneracy of $2N_F$.  This is the origin of
the discontinuity at $p=\mu$, since the Fermi momentum remains at this
point for these free quarks.  Thus the plot should be considered a combination
of a smooth distribution of Bose-condensed diquarks added to a simple
step function of magnitude 4.  To be explicit, \eq{fulldens} can be
rewritten as
\beqa
n^{(2)} &\pj=&\pj \frac{2 N_f(N_c-2)}{6\pi^2}\mu^3 \nonumber\\
&&+ 2 N_F \int\dbar{p} \,
\frac{\mu(p_4^2+\mu^2-p^2)}{[p_4^2+(p-\mu)^2][p_4^2+(p+\mu^2)^2]}
\left[\frac{1}{1+4\beta(p)\Delta^2} \right]\,.
\label{ph2dens}
\eeqa
The first term is the density for $N_c-2$ free quarks, the second
a contribution from the remaining degrees of freedom which are no longer
properly described as single quarks.

\begin{figure}[tb]
\setlength\epsfxsize{10cm}
\centerline{\epsfbox{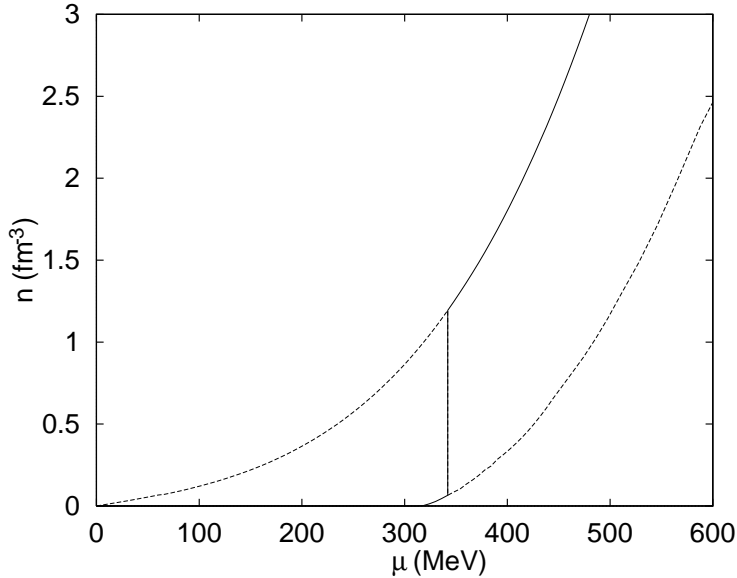}}
\caption{Quark density as a function of chemical potential.  The solid line
denotes the the equilibrium state of quark matter, the dashed lines the
metastable continuations.  Naive quark matter saturation density lies at
$\approx 0.45 \, {\rm fm}^{-3}$.}
\label{dens}
\end{figure}

Performing the integration over all momenta, we arrive at the density
as a function of chemical potential.  In \fig{dens} this is shown for
the equilibria states, demonstrating the large discontinuity at the
phase transition, where the dashed lines are the metastable continuations in
each phase.
Since the constituent mass of a quark in the chiral
breaking phase remains near 300 MeV, no density is amassed below this
chemical potential.  This is consistent with our finding a virtually
constant chiral condensate below the critical chemical potential, in that
the absence of physical quark density should leave the physical 
condensate unchanged from its vacuum value.  Soon after the point at 
which the Fermi surface moves from zero and physical states begin to be 
filled, the onset of the phase transition has been reached.  Thus, for 
pure quark matter in this simple mean-field treatment, the maximum 
density of a purely chiral-broken phase is a rather small $n = 
0.062\,{\rm fm}^{-3}$.  Taking a naive saturation density for quark 
matter as three times that for nuclear matter, we have $n_{0} = 0.445\, 
{\rm fm}^{-3}$, which
lies within the coexistence line between the two phases. It is only above
a density of $n = 1.05 \,{\rm fm}^{-3} = 2.53 n_{0}$ that one finds
the pure superconducing phase.  For reference, the density of free quarks
at this $\mu$ would be $2.25 n_{0}$, somewhat lower than that which
is found in the pairing scenario where the states are in some sense
compressed through bosonic condensation.

This extremely low density for the onset of chiral restoration suggests
that in the core of heavy nuclei quarks exist in some transitional,
`boiling' state \cite{Princeton/MIT}.
However, we have disregarded certain effects which could modify these
results.  Considering an extended mean-field ansatz of more than the
three channels we consider would likely result in additional interquark
forces, which could lead to enhanced repulsion and extend the stability of
the chiral-breaking phase to higher density.  We have also not taken
into account the binding energy of the quarks within a nucleon, which would
not modify our results at the quark level but could lead to some
differences between quark and nuclear matter.

Another possibility is to vary the parameters of our theory.
In particular, one can change the instanton packing fraction which has been
fixed at $\bar{\rho}/\bar{R} = 1/3$ in all calculations described here.
This in turn scales the strength of
the induced interaction.  However, taking values from half to twice our
choice does not lead to qualitative differences in the density results;
in every case chiral symmetry is restored at densities which are a small
fraction of that for saturation.

Finally, we note that the strong prediction one can make from this and
similar treatments is the presence of a first-order transition which
restores chiral symmetry.
It is a conclusion which is
shared by other recent works which consider chiral symmetry breaking at
the quark level.  In particular we note that the density discontinuity which
we find at this transition, $\Delta n \approx 2.4 n_{0}$, is close to that
found in the NJL-type treatment of Berges and Rajagopal \cite{BR}
($\Delta n \approx 1.5 n_0 - 2 n_0$) as well as that from the chiral
random matrix model ($\Delta n \approx 2.5 n_0$) \cite{HJSSV}.

\section{Conclusions}

We have formulated the effective low-energy action for two light fermions
induced by instantons at nonzero chemical potential.  In the resulting
fermion vertex we have retained the full dependence on both momentum
and chemical potential, which arises from the would-be zero modes.
The overall interaction strength is given by a dynamically-determined
coupling, which is a consequence of the fact that the instanton weight
itself is proportional to the fermion determinant.
In these respects we differ from other studies such as the random matrix
model, Nambu-Jona-Lasinio models, and alternative instanton approaches.

In general, introduction of the fermion chemical potential leads to
complications due to the resulting non-hermiticity of the Dirac operator.
This is particularly pronounced through the complex nature of the quark
interaction, induced by the zero-mode solution of the complex
eigenvalue problem.  By retaining the exact functional forms without
simplifications we are able to avoid any complications from individual
imaginary eigenvalues.  
This is an advantage over various numerical techniques,
and should persist for all orders in the instanton density.
The formalism outlined in this paper allows one 
to make such an expansion in a systematic way. 

The effective action leads to a competition between two phases, one of
chiral symmetry breaking and another characterized by diquark
condensation.  It was studied by solving a coupled system of gap
equations to first order in the instanton density.  For two massless
flavours and three colours, a simple ansatz allowed us a detailed study of
the various types of symmetry breaking.  Although we did recover an
interesting case of mixed condensation which broke both chiral and colour
symmetries, it was found to be thermodynamically disfavoured.  Considering
the remaining possibilities, we find that spontaneously broken chiral
symmetry is restored through a first order phase transition,
replaced by colour breaking due to the formation of a diquark
condensate.  With our `standard' choice of the instanton ensemble,
$N/V = 1\,{\rm fm}^{-4}$ and $\bar{\rho}/\bar{R} = 1/3$, we find the
critical chemical potential is $\simeq 340$ MeV,
where the superconducting energy gap is 115 MeV and decreasing with
rising $\mu$.
This translates into the onset of chiral
restoration at very low density of quark matter. Indeed,
the critical $\mu_c=340$ MeV corresponds to the quark density
of $0.14\;n_0$ in the chiral broken phase and to the density of
$2.53\;n_0$ in the superconducting phase, where $n_0=0.445\;{\rm fm}^3$
is the quark density corresponding to the standard nuclear matter.
Taken literally, it suggests that the deep
interior of heavy nuclei are in a `boiling' mixture of the two
phases, as suggested earlier in ref. \cite{Princeton/MIT}.
It should be kept in mind however, that our calculations were to only 
first order in the instanton density and that we neglected the
binding of constituent quarks into nucleons. Both circumstances 
might shift the critical density somewhat.  

Restricting our discussion to the quark level, we computed the fermion
occupation numbers in the two distinct phases which explicitly demonstrate
the physics in each system.  In the chiral broken phase, the reduced
Fermi radius illustrates the effective mass, whereas in the diquark phase
BCS-type behaviour is demonstrated by the lack of a sharp Fermi surface.

\subsection*{Acknowledgements}

One of us (G.W.C.) thanks A.D. Jackson for interesting and exceedingly
useful discussions.

\section*{APPENDIX. Fourier Transforms of Fermion Zero Modes}

The use of the exact fermion zero modes in the momentum space
tremendously simplifies all calculations. The starting point is
the exact fermion zero mode in the field of one (anti)instanton
in $x$ space \cite{Abr,DeCarv}, which we cite for arbitrary
instanton position $z$, size $\rho$ and orientation given by
rectangular $N_c\times 2$ matrix $U$. In the spinor representation
for the Dirac matrices the zero modes are two-component
Weyl spinors which can be written as:

\beq
\left[\Phi_{L,R}(x-z)\right]^{\alpha i} =
\frac{\rho}{\sqrt{2}\pi} e^{\mu (x_4-z_4)} \sqrt{\Pi(x-z)}
\partial_\mu
\left( \frac{e^{-\mu (x_4-z_4)}\Delta(x-z,\mu)}{\Pi(x-z)} \right)
\left(\sigma_\mu^\pm\right)^i_j\epsilon^{jk}U^\alpha_k\,.
\la{zeromode_x}\eeq
The single instanton solution is apparent in the functions
\beqa
\Pi(x) &\pj=&\pj 1 + \frac{\rho^2}{x_4^2+r^2+\rho^2} \; ,\nonumber\\
\Delta(x,\mu) &\pj=&\pj \frac{1}{x_4^2+r^2}\left[ \cos(\mu r) + \frac{x_4}{r}
\sin(\mu r) \right]\,.
\eeqa
The Fourier transform is defined as
\beq
\Phi(p,\mu) = i \int {\rm d}^4x\,e^{-ip\cdot x}\Phi(x,\mu)
\eeq
and has the structure
\beq
\Phi_{L,R}(p,\mu)^{\alpha i} = \varphi_\mu(p,\mu)
\left(\sigma^{\pm}_{\mu}\right)^i_j\epsilon^{jk} U^{\alpha}_k \,.
\eeq
Lorentz symmetry is broken at finite $\mu$, and the components
become
\beqa
\varphi_4(p_4,p;\mu) &\pj=&\pj \frac{\pi\rho^2}{4p} \Big\{
(p-\mu-ip_4)\left[(2p_4+i\mu)f_{1-} + i(p-\mu-ip_4)f_{2-}\right]
\nonumber\\
&&\quad\quad+(p+\mu+ip_4)\left[(2p_4+i\mu)f_{1+} -
i(p+\mu+ip_4)f_{2+}\right]\Big\}\,,
\nonumber\\
\varphi_i(p_4,p;\mu) &\pj=&\pj \frac{\pi\rho^2 p_i}{4p^2} \Bigg\{
(2p-\mu)(p-\mu-ip_4)f_{1-}+(2p+\mu)(p+\mu+ip_4)f_{1+} \nonumber\\
&&\quad\quad + \left[2(p-\mu)(p-\mu-ip_4) -
\frac{1}{p}(\mu+ip_4)[p_4^2+(p-\mu)^2] \right] f_{2-}
\nonumber\\
&&\quad\quad +\left[2(p+\mu)(p+\mu+ip_4) +
\frac{1}{p}(\mu+ip_4)[p_4^2+(p+\mu)^2] \right] f_{2+}\Bigg\}\,,
\eeqa
where the scalar $p = \vert\vec p\vert$, the spatial $i=1\dots 3$,
and the functions
\beq
f_{1\pm} = \frac{I_1(z_{\pm})K_0(z_{\pm}) - I_0(z_{\pm}) K_1(z_{\pm})}{z_{\pm}}
\,, \quad
f_{2\pm} = \frac{I_1(z_{\pm})K_1(z_{\pm})}{z_{\pm}^2}
\eeq
are evaluated at $z_{\pm} = \frac{1}{2}\rho\sqrt{p_4^2+(p\pm\mu)^2}$.
With these expressions it is explicitly verified that the normalization
condition holds for any $\mu$:
\beq
1 =\int\!\frac{d^4p}{(2\pi)^4}\:\tilde\Phi_I(p,\mu)\Phi_I(p,\mu)=
\int\!\frac{d^4p}{(2\pi)^4}\:\left[ \varphi_4^*(-\mu)\varphi_4(\mu)
+ \vec\varphi^*(-\mu)\cdot\vec\varphi(\mu)\right].
\la{1}\eeq

\end{document}